\documentclass[%
 reprint,
nofootinbib,
 amsmath,amssymb,
 aps,
]{revtex4-2}

\usepackage{graphicx}
\usepackage{dcolumn}
\usepackage[shortlabels]{enumitem}

\usepackage{comment}

\newcommand{\mycomment}[1]{}

\newcommand{\be}{\begin{equation}}
\newcommand{\ee}{\end{equation}}

\usepackage{soul,xcolor}
\setstcolor{red}

\begin{document}

\title{Universal black hole solutions for all F(R) gravitational theories}%
 
\author{Cristobal Laporte}
\email{cristobal.laportemunoz@ru.nl}
\author{Agustín Silva}
\email{agustin.silva@ru.nl}
\affiliation{High Energy Physics Department, Institute for Mathematics, Astrophysics, and Particle Physics, Radboud University, Nijmegen, The Netherlands.}

\date{\today}

\begin{abstract}
Extended gravitational models have gained large attention in the last couple of decades. In this work, we examine the solution space of vacuum, static, and spherically symmetric spacetimes within $F(R)$ theories, introducing novel methods that reduce the vacuum equations to a single second-order equation. For the first time, we derive analytic expressions for the metric functions in terms of the arbitrary functional $F(R)$, providing detailed insight into how the gravitational action impacts the structure of spacetime. We analyze conditions under which solutions are asymptotically flat, regular at the core, and contain an event horizon, obtaining explicit expressions for entropy, temperature, and specific heat in terms of $F(R)$. By using a single metric degree of freedom, we identify the most general solution and examine its (un)physical properties, showing that resolving singularities is not possible within this restricted framework in vacuum. For the general case involving two metric functions, we use several approximation schemes to explore corrections to Schwarzschild-(anti)de Sitter spacetimes, finding that $F(R)$ extensions to General Relativity induce instabilities that are not negligible. Finally, through an analysis of axial perturbations, we derived a general expression for the potential of quasinormal modes of a black hole as a function of the arbitrary Lagrangian.
\end{abstract}

\maketitle

\section{Introduction}\label{sect.1}

General relativity (GR) remains the most successful framework for describing gravity, accurately predicting gravitational effects on both solar system and cosmological scales. Such predictions align remarkably well with observations, further corroborated by the recent detection of gravitational waves from binary black hole mergers \cite{LIGOScientific:2016aoc,KAGRA:2021duu} and the shadow imaging of supermassive black holes \cite{EventHorizonTelescope:2019dse,EventHorizonTelescope:2019ggy,EventHorizonTelescope:2022wkp}, both of which are consistent with classical GR. While this agreement is robust, some tensions have emerged over the past few decades, such as the $H_0$ \cite{Verde:2019ivm} and $\sigma_8$ \cite{Abdalla:2022yfr} tensions. Furthermore, theoretical conflicts between quantum mechanics and GR, including the prediction of spacetime singularities and a lack of explanation for the origin of dark matter and dark energy, strongly suggest that GR alone may not provide a complete description of gravity.

The preceding arguments suggest that GR is a low-energy effective approximation of a more complete theory. When the effective action is written just in terms of the metric field, higher curvature corrections to the Einstein-Hilbert term arise \cite{Donoghue:1994dn,Burgess:2003jk}. This results from either the introduction of new classical high-energy degrees of freedom or from quantum corrections upon integrating out massive states. Although higher-order terms are often presumed negligible in regions of low curvature, in more general contexts the solution space exhibits greater complexity compared to classical scenarios \cite{Lu:2015cqa,Bueno:2017sui}. Exploring these alternatives is theoretically well-motivated and yields intriguing outcomes. However, the inclusion of higher-order corrections complicates the equations of motion, leading to higher-order differential equations and often resulting in Ostrogradsky instabilities \cite{Stelle:1977ry,Woodard:2015zca}, which manifest themselves as massive ghosts that violate unitarity.

Among the numerous modifications to GR, $F(R)$ gravity is one of the most widely studied and conceptually simple models, where the Lagrangian density is an arbitrary function of the Ricci scalar $R$ \cite{Sotiriou:2008rp,DeFelice:2010aj}. As a defining feature, this $F(R)$ theory aligns with a generalized Brans-Dicke theory, incorporating a scalar potential where a scalar field is non-minimally coupled to gravity \cite{Faraoni:2010pgm}. Consequently, $F(R)$ models naturally evade the Ostrogradsky instability \cite{Woodard:2006nt}, as all higher-derivative terms can be encompassed into the definition of the scalar field. This property makes $F(R)$ theories particularly valuable for describing early-universe inflation \cite{Starobinsky:1980te} and late-time cosmic acceleration \cite{Nojiri:2003ft,Nojiri:2010wj}.

Despite their success in explaining cosmological observations, studies on static and spherically symmetric (SSS) vacuum solutions within $F(R)$ theories have been restricted to specific cases in the Lagrangian structure, due to the difficulties in obtaining exact solutions to the fourth-order field equations in terms of the metric components. Consequently, the solution space has not been fully explored. SSS solutions have been limited by certain simplifying assumptions, which we list below:
\begin{enumerate}[(a)]
    \item Due to the complexity of the equations of motion in a general $F(R)$ theory, many studies have focused their attention on the constant curvature $R$ regime \cite{Multamaki:2006zb,delaCruz-Dombriz:2009pzc,delaCruz-Dombriz:2012bni,Casado-Turrion:2023rni}. In the Einstein-Hilbert theory, this approach provides the most general SSS solution; however, such generality is not guaranteed for an arbitrary $F(R)$. Finding SSS solutions without imposing the constant curvature condition is challenging, and only perturbative solutions have been obtained so far.
    \item To streamline the analysis, numerous studies on vacuum solutions within general $F(R)$ Lagrangians have been conducted under the assumption of a single metric function \cite{Mo:2016apo,Tang:2020sjs,Karakasis:2021lnq}. While describing spacetime with one metric function can greatly simplify the study of thermodynamic and structural aspects, such as the existence of hairy solutions \cite{Canate:2015dda} consistent with the null energy condition, this approach limits the solution space and may omit essential mathematical and physical features. For instance, asymptotic flatness in the absence of a cosmological constant has not been investigated in detail under this simplification. Furthermore, in some scenarios, only one component of the equations of motion is analyzed, naively assuming that the system admits such single metric function solutions.
    \item Although the equations of motion and certain general properties, such as entropy, are well understood, much of the existing literature examines specific forms of $F(R)$ to analyze the spacetime structure, the behavior of metric functions, and thermodynamic properties across different regions \cite{Nashed:2020mnp,Elizalde:2020icc,Nashed:2021lzq}. It would, therefore, be highly desirable to derive physical properties for a general $F(R)$. Such an approach would facilitate comparisons between popular cosmological models and black hole-like solutions, offering a more complete perspective on their compatibility or potential discrepancies.
\end{enumerate}
In this work, we take the first steps to address these challenges from a general perspective. We introduce a novel method to reduce the order of the vacuum equations of motion, enabling a critical assessment of the common assumptions listed above, and at the same time, deriving conditions on the $F(R)$ Lagrangians that are compatible with fundamental requirements, such as asymptotic flatness and the absence of curvature singularities throughout the spacetime.

The paper is organized as follows. Section \ref{sect.2} presents the field equations of motion for SSS spacetimes, where we reduce the differential equation order from fourth to second by replacing one of the unknown metric functions with the scalar curvature. We solve analytically the metric functions in terms of the scalar curvature and an arbitrary Lagrangian $F(R)$, and we demonstrate that the complete information of the system can be captured by a single non-linear, second-order differential equation involving only the scalar curvature. Section \ref{sect.3} specifies the conditions that must be imposed on $F(R)$ and the metric functions to ensure asymptotic flatness, horizon existence, and core regularity. In Section \ref{sect.thermodynamics}, we analyze the thermodynamic properties of black holes in $F(R)$ theories, deriving expressions for temperature, entropy, and specific heat. In Section \ref{sect.6}, we critically assess the implications of using a single metric function to describe spacetime. Section \ref{sect.approximations} focuses on different approximation schemes to treat the case of SSS spacetimes with $2$ metric degrees of freedom, and we study generalizations of the Schwarzschild-(anti)de Sitter case both analytically and numerically. In section \ref{sec:gravgraves} we derive a general expression for the potential of quasinormal modes in $F(R)$, by performing axial perturbations of the obtained analytical solutions. We conclude with a summary of our findings and discuss potential future directions.

\bigskip


\section{Equations of motion and Exact solutions}\label{sect.2}

\bigskip


We use the action 
\be\label{ActionI}
\Gamma = \frac{1}{{16 \pi  G^2}}\int dx^{4} \sqrt{-g}  \left(G R+F(G R)\right) \, ,
\ee
where $R$ is the Ricci scalar, $g$ the metric determinant, $G$ Newton's constant, and $F$ is some dimensionless scalar function. Throughout this work, we make $G=1$, absorbing it into the coordinates.
The vacuum field equations of \eqref{ActionI} are derived from the variational principle and are given by, 
\be 
    R_{\mu\nu} - \frac{g_{\mu\nu}}{2}  \, R + R_{\mu\nu} F_{R} - \frac{g_{\mu\nu}}{2}  \, F + \left(\nabla_{\mu} \nabla_{\nu} - g_{\mu\nu} \Box \right) \, F_R = 0.
    \label{eq:vacuumeqfr}
\ee
Here, $R_{\mu\nu}$ denotes the Ricci tensor, $\Box=\nabla_{\mu} \nabla^{\mu}$ is the d'Alambertian operator and $F_R \equiv \frac{dF}{dR}$. Derivatives of $F$ of order $n$ with respect to $R$ will be denoted by the addition of $n$ times the subscript $_{R}$. In this work, we will study SSS solutions of the vacuum field equations \eqref{eq:vacuumeqfr}. These types of solutions can be found by inserting the ansatz for the metric 
\be
 ds^{2}=-B(r)e^{\varphi(r)}dt^{2}+\frac{1}{B(r)}dr^{2}+r^{2}d\Omega^{2} \, ,
 \label{eq:ansatzmetric}
\ee
which leads to two independent equations of motion for $B(r)$ and $\varphi(r)$. The angular component of the metric is related to these equations through the radial component of the Bianchi identity.

If one works directly with the metric degrees of freedom ($B, \, \varphi$) as the unknown functions of the system, and uses the tt and rr-components of \eqref{eq:vacuumeqfr}, then the two equations will form a system of fourth order, non-linear, coupled differential equations (see equations \eqref{eq:tteqvacuum} and \eqref{eq:rreqvacuum} in the appendix). Solving that system directly, either analytically or numerically, can be challenging. In this work, we adopt a different approach to simplify the system.

To find analytic solutions to this system of equations, it is convenient to perform a change of variables and trade one of the metric functions by the Ricci scalar $R$. Typically, the Ricci scalar is derived from the metric functions $B$ and $\varphi$, once their solutions are determined, and computes
\be
 R=\frac{1}{2 r^2} [4-2 r^2 B''-r B' (3 r \varphi
   '+8)-B (2 r^2 \varphi ''+(r \varphi
   '+2)^2)] ,
\label{eq:ricciscalarfrommetricfunctions}
\ee
where from now on the primes will denote derivatives with respect to $r$.

Given that the Lagrangian of the theory is conveniently expressed in terms of the Ricci scalar, as are the equations of motion, we will exploit this feature to trade the independent variables of the system to be solved. Specifically, rather than solving the vacuum equations for $(B, \, \varphi)$ and later finding $R$ as a function of the two metric functions using \eqref{eq:ricciscalarfrommetricfunctions}, we will solve the system for $R$ and one of the metric variables, say $B$. By doing this, one is simply changing the unknown variables from $(B, \, \varphi)$ to $(B, \, R)$. Naturally, this requires determining the exact relation of $\varphi$ as a function of $R$ and/or $B$. 

We will show how this transformation not only enables us to solve the exact dependence of $\varphi$ as a function of $R$ and $B$, but it also allows us to find the exact dependence of $B$ as a function of $R$, leaving a single equation for $R$ to be solved. In order to do this, we use linear combinations of the tt and rr-components of \eqref{eq:vacuumeqfr} (\eqref{eq:tteqvacuum}, \eqref{eq:rreqvacuum}) together with \eqref{eq:ricciscalarfrommetricfunctions}. As mentioned before, the goal is to reduce the system to one in which $B$ and $\varphi$ are expressed in terms of $R$, leaving a single equation for $R$ to be solved. The result of this linear combination is

\begin{widetext}

\begin{align}\label{eq:R}
    R'' = \frac{1}{6 r B \, F_{R} \left(1 + F_R\right)} \Big(& 6 R' \, F_{R} \left[r \, B \, R' \, F_{R} - \left(1 + B\right) \, \left(1 + F_{R}\right)\right] + r F \left(4 + 4 F_R - r \, R' \, F_{R}\right) \nonumber\\
    &+ r R \left[2 - 2 F^2_{R} + r \left(1 + 2 F_R\right) R' \, F_{R}\right] - 6 r \left(1 + F_R\right) B \, {R'}^2 \, F_{RR} \Big) \, ,
\end{align}

\begin{align}\label{eq:B}
    B' = \frac{1}{3 r \left(1 + F_R\right) \left(2 + 2 F_R + r R' \, F_{R}\right)} \Big(&r R' \, F_{R} \left[6 \left(1 + F_R \right) + r^2 \left(F - R \left(1 + 2 F_{R}\right)\right) - 6 B \left(1 + F_{R} + r R' \, F_{R}\right)\right] \nonumber\\
    &+\left(1 + F_R\right) \left(6 \left[1 + F_R - B \left(1 + F_R\right)\right] - r^2 \left[F + R \left(2 + F_R\right)\right]\right)\Big) \, ,
\end{align}

\begin{align}\label{eq:varphi}
    \varphi' = \frac{1}{3 B \left(1 + F_R\right) \left(2 + 2 F_R + r R' \, F_{R}\right)} \Big(&2 r \left(1 + F_R\right) \left(2 F + R - R \, F_R\right) + 6 r \, B \, {R'}^2 F^2_{R}\nonumber\\
    &-\left[6 + 6 B + r^2 \left(F - R\right) + 2 \left(3 + 3 B - r^2 R\right) F_R \right] \, R' \, F_{R}\Big) \, .
\end{align}

\end{widetext}

The obtained system \eqref{eq:R}-\eqref{eq:varphi} has only $2$ independent functions, as in the original setup (\eqref{eq:tteqvacuum}, \eqref{eq:rreqvacuum}). Several implications arise from this formulation. First, the original system of $2$ equations of fourth order, has been transformed into a system of $3$ equations of at most second order, and where the highest derivative can be isolated on one side of the equations, making the system amenable to numerical solvers. Second, by treating $R$ as an independent variable, both $B$ and $\varphi$ can be solved analytically in terms of $R$ for any given Lagrangian $F$.

 Even though \eqref{eq:varphi} can simply be solved by integrating both sides, one can further simplify the solution by replacing $B$ from \eqref{eq:R} into \eqref{eq:varphi}. This leads to a very simple expression for $\varphi$
\be
    \varphi(r) = \varphi_0 + \int dr \, \frac{2 r \left(F_{R} \, R'' + {R'}^2 \, F_{RR}\right)}{2 + 2 F_{R} + r R' \, F_{R}} \, ,
    \label{eq:phisolanalytic}
\ee
where $\varphi_0$ is a free constant, present due to the time re-parameterization invariance of the metric.

Note that (\ref{eq:B}) is independent of $\varphi$, and it is a linear first order ODE, that also admits an analytical solution given by 

\be
 B(r)= \frac{e^{\int dr \, Q(r)}}{r
   \left(1+F_{R}\right)^2}  \left(c_1+\int dr \, \frac{e^{-\int dr \,Q(r)}\, P(r)}{(1+F_{R}+ \frac{1}{2} r R'
   F_{R})}
    \right),
    \label{eq:Bsolanalytic}
\ee
 with $c_1$ an arbitrary constant,
\be
Q(r)=\frac{2}{r} \left(\frac{3}{2}-\frac{3 \left(1+F_{R}\right)}{2+2 F_{R}+r R'
   F_{R}}\right), 
\ee
and 
\begin{align}
    P(r)=&\frac{\left(1+F_{R}\right)^2}{6} [\frac{r R' \left(r^2 \left(F-R \left(1+2F_{R}\right)\right)\right) F_{R}}{\left(1+F_{R}\right)}- \nonumber \\
    &  \left(r^2
   \left(R \left(2+F_{R}\right)+F\right)-6 \left(r R'F_{R}+1+F_{R}\right)\right)].
\end{align}
 
With these exact solutions for the metric functions in terms of the now independent variable $R$, the only remaining equation to be solved is \eqref{eq:R}. This significantly reduces the complexity of the original system. First, this allows us to directly see the impact of different Lagrangians on the metric. Second, this enables us to perform calculations using the found solutions, propagating the arbitrary $F(R)$ to see its impact on physical quantities, before finally solving the equation for $R(r)$. 

Moreover, from solutions \eqref{eq:phisolanalytic} and \eqref{eq:Bsolanalytic}, along with equation \eqref{eq:R}, one can see that for a given arbitrary Lagrangian $F(R)$, the solution space for SSS spacetimes is completely parameterized by, at the most, four arbitrary constants. Two of them can already be seen in \eqref{eq:phisolanalytic} and \eqref{eq:Bsolanalytic}, one for each metric function, due to the first-order nature of the equations \eqref{eq:B} and \eqref{eq:varphi}. The remaining two constants arise from solving \eqref{eq:R}, which is a second-order equation. This constitutes a major reduction compared with the standard analysis of the system of coupled fourth-order equations directly derived from \eqref{eq:vacuumeqfr}. Any local study of the solution space in $F(R)$ approaches, such as a Frobenius analysis, should take these results into account. 

Our results also facilitate an analysis of how different boundary conditions, such as the imposition of an event horizon, or regularity at the origin affect the solutions, as we will see in the next sections. Furthermore, a wide range of approximations and assumptions can be applied to these solutions, depending on the form of the arbitrary function $F(R)$. These include low-curvature expansions, near-extremal expansions, and other approaches that will be discussed in subsequent sections.

\subsection*{\textit{The case of constant scalar curvature}}\label{sec:constantcurvature}

One can check that our solutions recover the well-known case of the Schwarzschild-(anti)de Sitter solution when one forces the scalar curvature to be a constant $R=R_0$. Indeed, in this case \eqref{eq:R} reduces to 
\be
 F(R_0)=\frac{R_0}{2}(F_R(R_0)-1) \, ,
 \label{eq:constricciscalareq}
\ee
and the solutions \eqref{eq:phisolanalytic} and \eqref{eq:Bsolanalytic} for $B(r)$ and $\varphi(r)$ simplify to
\be
B(r)=1 + \frac{c_1}{r} - \frac{R_0}{12}r^2 \,\,\,\,\,\,\, \varphi(r)=\varphi_0 \, .
\label{eq:constricciscalarmetricfunctions}
\ee 

This is the Schwarzschild-de Sitter solution. It is important to note that obtaining this solution does not require $F(R_0)=-2\Lambda$ and $R_0=4\Lambda$ as in the Einstein-Hilbert action. Instead, as long as $R_0$ is a solution of \eqref{eq:constricciscalareq}, Schwarzschild-de Sitter is the only possible SSS solution for constant scalar curvature. Our results demonstrate that, in the $F(R)$ framework, there are no alternatives to the Schwarzschild-(anti)de Sitter solution when constant curvature solutions are considered. This result is not new, and was only included as a consistency check in this work. The case of constant scalar curvature was extensively studied in \cite{delaCruz-Dombriz:2009pzc}, and therefore we will not explore it further in this work.

\section{Asymptotically flat, black hole and regular spacetimes}\label{sect.3}

In this section, we derive some general properties that must be satisfied by the SSS solutions coming from the action in \eqref{ActionI}, based on the governing equations \eqref{eq:R}–\eqref{eq:varphi}. While obtaining exact analytic solutions for \eqref{eq:R} is not feasible in the most general case, we can still examine important qualitative features by analyzing the behavior of the metric solutions \eqref{eq:phisolanalytic} and \eqref{eq:Bsolanalytic} in different spacetime regimes.

Specifically, we investigate the constraints imposed on the metric functions to ensure regularity at $r=0$, as well as the conditions necessary for the existence of an event horizon. In addition, we linearize the equations of motion around flat spacetime to obtain general expressions for the perturbations around Minkowski spacetime, which are written in terms of the function $F$ and its derivatives.

\subsection{\textit{Corrections in the asymptotic region}}\label{sec:asymptoticsol}

Our first step is to analyze the asymptotic degrees of freedom that come from the higher derivative terms in the action, and capture their impact on the asymptotic limit ($r \to \infty$) of the metric.
To do that, we make use of the fact that to recover asymptotically flat spacetime, the scalar curvature must be $0$ at infinity, and therefore we assume that
\be 
    R(r) =  \, \epsilon(r) \, , \quad |\epsilon(r)| \ll 1 \, ,
\ee
and expand solutions \eqref{eq:phisolanalytic} and \eqref{eq:Bsolanalytic}, and equation \eqref{eq:R} to linear order in $\epsilon(r)$. We will neglect the cosmological constant by setting $F(0)=0$, and we assume that the linear term in the action is completely captured by Einstein's linear term, which allows us to set $F_{R}(0)=0$. Under these assumptions, working at first order in the perturbation $\epsilon(r)$, one obtains from \eqref{eq:R} that
\be
    \epsilon(r) = \frac{1}{r}\left(c_1 e^{-m_0 \, r}  +  c_2 e^{m_0 \, r} \right) \, ,
    \label{eq:asymptoticsolRflat}
\ee
with ${m_0}^2 \equiv (3 F_{RR}(0))^{-1}$, and $c_1$ and $c_2$ arbitrary constants. Replacing this into \eqref{eq:phisolanalytic} and \eqref{eq:Bsolanalytic}, one gets the asymptotic solutions for the metric functions
\be
    \varphi(r) = \text{$\varphi_0 $} - \frac{c_1 \, e^{- m_0 \, r}}{3 m_0^2 r} (2 + m_0 \, r)+ \frac{c_2 \, e^{ m_0 \, r}}{3 m_0^2 r} (2 - m_0 \, r) \, ,
    \label{eq:asymptoticsolphi}
\ee
\be
    B(r) = 1 - \frac{2 M}{r} + \frac{c_1 \, e^{- m_0 \, r}}{3 m_0^2 r} (1 + m_0 \, r)+ \frac{c_2 \, e^{ m_0 \, r}}{3 m_0^2 r} (1 - m_0 \, r) \, ,
    \label{eq:asymptoticsolB}
\ee
where the arbitrary constant in \eqref{eq:Bsolanalytic} was renamed $-2M$.

Our linearized analysis reveals an additional degree of freedom beyond the standard transverse and traceless graviton propagating modes in vacuum \cite{Barausse:2008xv,Myung:2016zdl}, with a mass given by ${m_0}^2 = (3 F_{RR}(0))^{-1}$. Note that for $F_{RR}(0) < 0$, the mass $m_0$ becomes purely imaginary $m_0 = i \bar{m}_0$. In this case, the exponential corrections to Schwarzschild become oscillatory functions, and the presence of non-suppressed oscillatory terms appears when $c_1 \neq 0$ and/or $c_2 \neq 0$. In this case, these non-decaying oscillations prevent the metric from being asymptotically flat as $r \to \infty$, unless $c_1 = c_2 = 0$, i.e., the metric functions are exactly equal to the Schwarzschild metric. The only way to obtain non-trivial solutions while preserving asymptotic flatness at spatial infinity is to impose $0\leq F_{RR}(0)$, and thus we will restrict our analysis only to this case in the rest of this work.

One can also notice that if $F_{RR}(0) = 0$, then there are no asymptotically flat first-order corrections to the Schwarzschild solution. We will not focus on this situation in this paper, and leave its analysis for future work. 

\subsection{\textit{General solutions with a horizon and regularity conditions}}

Now we move to the interesting case of solutions with event horizons, which can be obtained in two distinct ways. The first and also the most straightforward approach, is to force $B(r_h)=0$ at some value $r=r_h$, if $r_h>0$, by appropriately fixing the arbitrary constant that appears in \eqref{eq:Bsolanalytic}. For completeness, the solution found by imposing a horizon at $r=r_h$ using this method is
\be
 B_{h}(r)= \frac{e^{\int_{r_h}^{r} dy \,Q(y)}}{r
   \left(1+F_{R}\right)^2}  \int_{r_h}^r dx \, \frac{e^{-\int_{r_h}^{x} dy \,Q(y)}\, P(x)}{(1+F_{R}+ \frac{1}{2} x R'
   F_{R})}
    \, ,
    \label{eq:Bsolanalytichorizon}
\ee
and 
\be
    \varphi_{h}(r) = \varphi_h + \int_{r_h}^{r} dx \, \frac{2 x \left(F_{R} \, R'' + {R'}^2 \, F_{RR}\right)}{2 + 2 F_{R} + x R' \, F_{R}} \, , 
    \label{eq:phisolanalytichorizon}
\ee
with $\varphi_h$ the (arbitrary) value of $\varphi$ at the horizon.

Notice that these solutions do not guarantee regularity at the origin, nor asymptotic flatness; these conditions must be verified for any specific choice of $F$. 

A second method to impose the existence of a point $r_h$ where $B(r_h)=0$ comes from noticing that $B(r)$ appears linearly on equation \eqref{eq:R}. This allows us to impose a local condition on the scalar curvature to guarantee the presence of a zero in $B$. In particular, from \eqref{eq:R} one can notice that if at some point $r_h$ the following conditions hold simultaneously
\begin{align}
 [-6 & r R'' \left(F_R+1\right) F_{R}-6
   R' \left(F_R+1\right) F_{R} \nonumber \\ 
  &  +6 r
   R'^2 \left(F_{R}^2-F_{RR}
   \left(F_R+1\right)\right)]_{|_{r=r_h}}\neq 0 \, ,
   \label{eq:localconditionRPrimeHorizon1}
\end{align}
\begin{align}
   [ R' &\left(r^2  \left(F-R \left(2
   F_R+1\right)\right)+6 \left(F_R+1\right)\right) F_{R} \nonumber \\ 
   &-2 r \left(F_R+1\right)
   \left(-R F_R+2 F+R\right)]_{|_{r=r_h}}=0 \, ,
   \label{eq:localconditionRPrimeHorizon2}
\end{align}
then necessarily $B(r_h)=0$. This is independent of the fixing of the arbitrary constant $c_1$ appearing in \eqref{eq:Bsolanalytic}.

In both methods used to impose a zero of the metric function $B$, one cannot directly guarantee whether this zero corresponds to a Cauchy horizon or an event horizon. Determining which type of horizon is being enforced depends on the sign of the first derivative of the metric, given in both cases by 
\be
B'(r_h)=[\frac{r (F+R)}{6
   \left(F_R+1\right)}-\frac{1}{3} r
   R+\frac{1}{r}]_{|_{r=r_h}} \, .
   \label{eq:Bfisrtderhorizon}
\ee

In the case where the factor multiplying $R'$ in \eqref{eq:localconditionRPrimeHorizon2} is different from zero, the value of the scalar curvature at the horizon ($R_h$) is a free finite parameter. This can be seen by performing a local analysis of the equations, suggesting that $R_h$ could be adjusted to satisfy the previous conditions. In contrast, when this factor vanishes, such as when $F_{R}$ is zero at the horizon, then $R_h$ is fixed by \eqref{eq:localconditionRPrimeHorizon2}. One example of this scenario is when $F(R)=-2\Lambda $ (Einstein - Hilbert action). 

Besides the existence of a horizon, one might also wonder about the possibility of obtaining non-trivial solutions that are regular at $r=0$.
Expanding the complete basis of seventeen non-derivative curvature invariants, referred to as Zakhary-McIntosh invariants \cite{Zakhary:1997xas} in powers of $r$ near $r=0$, one can see that in order to avoid divergences the metric functions must only satisfy the conditions
\be
B(0)=1 \,\,\,\,\,\, B'(0)=0 \,\,\,\,\,\,\varphi'(0)=0 \, .
\ee

There is one simple way to obtain a non-trivial solution such that this happens, and that is simply setting $r_h=0$ on the solutions \eqref{eq:Bsolanalytichorizon} and \eqref{eq:phisolanalytichorizon}. For non-constant curvature solutions and non-trivial $F(R)$ Lagrangians, this can yield solutions that differ from Minkowski or (anti)de Sitter. 
On top of that, depending on $F$, it is possible to adjust the arbitrary constant in \eqref{eq:Bsolanalytic}, so that the metric is regular at the origin, and simultaneously impose condition \eqref{eq:localconditionRPrimeHorizon2}, choosing $R_h$ so that \eqref{eq:localconditionRPrimeHorizon1} is satisfied, and such that \eqref{eq:Bfisrtderhorizon} is greater than $0$. 

While this would lead to solutions that are both regular at the origin and possess an event horizon \cite{Bonanno:2022jjp}, by fixing these constraints one has exhausted the number of boundary conditions for the resolution of \eqref{eq:R}. This means that the asymptotic behavior will not be controlled. The recovery of Minkowski or (anti)de Sitter spacetime at asymptotic infinity will be subject to the parameters and the functional form of $F$.

A detailed exploration of this situation will be exposed in the second part of this work \cite{secondpartpapaer}, where multiple examples of $F$ will be contrasted with such constraints.

\smallskip


\section{Black hole thermodynamics}
\label{sect.thermodynamics}

In this section, we explore the thermodynamics of solutions with an event horizon. The existence of asymptotically flat black hole solutions with a horizon provides a way to precisely establish the relationship between the surface gravity $\kappa$ and the temperature, given by $T=\frac{\kappa}{2 \pi}$. This formula remains valid across various modified theories of gravity due to its robust geometric origin \cite{Gibbons:1976ue}. In the case of generic $F(R)$ theories, one can show using \eqref{eq:Bsolanalytichorizon} and \eqref{eq:phisolanalytichorizon}, that the temperature of the horizon becomes, 
\be\label{Temperature}
    T_{H} = \frac{e^{\frac{\varphi_h}{2}}}{4 \pi r_h} \, \Big| 1 + \frac{r^2_h \left[F(R_h) - R_h \, (1 + 2 F_R(R_h))\right]}{6(1 + F_R(R_h))} \Big| \, .
\ee
 Setting the $F(R)$ function to zero for all values of $R$, the temperature from GR black holes $T_{GR}=(4 \pi \, r_h)^{-1}$ is recovered from \eqref{Temperature}. Note that, in our more general case the temperature \eqref{Temperature} vanishes when the following condition 
\be\label{eq:ExtremalCondition}
    F_R(R_h) = \frac{6 + r_h^2 (F(R_h) - R_h)}{2 r_h^{\rm 2} R_h - 6} \, ,
\ee
is met. To the best of our knowledge, this is the first time that zero temperature black holes, from now on called ``extremal'', are identified in the context of $F(R)$ theories of gravity. One could attempt to consider \eqref{eq:ExtremalCondition} as a differential equation for $F(R)$, and obtain Lagrangians whose event horizon solution are always extremal. If one follows this path, the solution is $F_e(R)=C \left(-\sqrt{6-2 R r_h^2}\right)-R$, with $C$ an arbitrary constant. If one forces the first terms of the Taylor expansion in powers of $R$ of $F_e(R)$ to match those of GR, one must have $C = \sqrt{\frac{2}{3}} \Lambda$ and $r_h = \sqrt{\frac{3}{\Lambda }}$. This means that the only possible 'extremal' $F(R)$ that recovers GR at low scalar curvature is  $F_e(R)=-R-2 \Lambda  \sqrt{1-\frac{R}{\Lambda }}$. In this scenario, the scalar curvature cannot be larger than $\Lambda$, and the only extremal solutions correspond to solutions with a horizon radius $r_h$ equal to the cosmological horizon. We don't expect that the exploration of such extremal Lagrangian would prove fruitful in physical situations. 

Continuing with the discussion about thermodynamics, the notion of black hole's thermal radiation \cite{Hawking:1975vcx}, facilitates the determination of black hole entropy in any local covariant theory. In GR, black hole entropy is given by the Bekenstein-Hawking area law, $S_{GR}=\frac{A}{4}$, where $A$ denotes the area of the horizon. However, this entropy formula requires modifications when higher-order curvature terms are included in the gravitational theory. A generalized expression for black hole entropy is provided by the Wald entropy formula \cite{Wald:1993nt,Iyer:1994ys}, which allows entropy to be computed in a wide range of gravitational theories. Wald's formula reads
\be\label{WaldEntFor}
    S = -\frac{1}{8} \int_+ d\Sigma \, \sqrt{h} \frac{\delta L}{\delta R^{\mu\nu\rho\sigma}} \, \epsilon^{\mu\nu} \epsilon^{\rho\sigma} \, ,
\ee
where the integration is taken over the bifurcation surface of the horizon, $h$ is the determinant of the metric on the horizon, and $\epsilon^{\mu\nu}$ is the antisymmetric binormal tensor to the bifurcation surface, normalized such that $\epsilon^{\mu\nu} \epsilon_{\mu\nu} = -2$. Substituting our results from \eqref{eq:R}-\eqref{eq:varphi} into the Wald formula \eqref{WaldEntFor} gives 
\be\label{EntropyFR}
    S = \frac{A}{4} \left(1 + F_R (R_h)\right) \, .
\ee
This result is consistent with the entropy formula found in \cite{Cognola:2005de,Faraoni:2010yi}. To prevent negative values of the entropy, the condition 
\be 
    [\text{sgn}\left(2 F_R + r R' F_{R} \right) = \text{sgn}\left(F_R \, r^2 + (2-r^2 R) \, F_R \right)]_{|_{r=r_h}}
\ee
has to be satisfied. Notably, the condition for vanishing entropy, $F_{R}(R_h) = -1$, is not inherently incompatible with the extremal black holes condition \eqref{eq:ExtremalCondition}. Nevertheless, it could potentially lead to singularities on the solutions \eqref{eq:Bsolanalytichorizon} and \eqref{eq:phisolanalytichorizon}. It would be particularly interesting to investigate whether these conditions lead to singularities in the metric functions, or if consistent solutions can be found at the horizon. In the former case, the appearance of singularities suggests that SSS black hole solutions in $F(R)$ gravity might exclude extremal black holes with vanishing entropy, thus ruling out configurations where $T=0$ and $S=0$ simultaneously. In the latter case, if consistent solutions are found, the dynamics of this extremal scenario could place significant constraints on the parameter space of $F(R)$ theories, thereby limiting the freedom to construct viable black hole solutions. Further investigation in this direction could provide deeper insights into the nature of extremal black holes and their role in modified gravity theories. We expect to come back to this point in a future work.

To close this section, we compute the specific heat, defined as 
\be\label{eq:CvI}
    C_V = T \, \left(\frac{\partial S}{\partial T}\right)_M \, .
\ee
Parameterizing the specific heat in terms of the scalar curvature at the horizon $R_h$ and the position of the horizon $r_h$, $C_V$ can be written as 
\be\label{eq:CvFR}
    C_V =  \frac{2 \pi r^{\rm 2} \left(1 + F_R\right) \left[r^{\rm 2} (R - F) + 2 F_R (r^{\rm 2} R - 3)\right]}{6 + r^{\rm 2} (R - F) + 2 F_R \, (3 + r^{\rm 2} R)}\Big|_{r=r_h} \, .
\ee

The sign of the specific heat is especially important for determining the thermodynamic stability of black holes. A detailed analysis of its signature, and therefore the thermodynamic stability of black holes for various $F(R)$ gravitational models will be the subject of future work \cite{secondpartpapaer}.

\smallskip

\section{ The case of $\varphi = 0 $}\label{sect.6}

To explore a situation where the equation for $R(r)$ can be solved analytically, let's analyze the simplest possible case: when there is only a single metric degree of freedom. That is, when $\varphi(r)=0$. From \eqref{eq:phisolanalytic} we see that the only scenario in which this happens is when 
\be
F =-2\Lambda + \int\,dr\, (\alpha\, r+\beta )R'(r) \, ,
\label{eq:phi0constraint}
\ee
where $\alpha$ and $\beta$ are arbitrary constants, with the only possible restriction being that $\alpha\, r+\beta  \neq -1$ for all $0\leq r$, or otherwise singularities could appear throughout the equations and solution \eqref{eq:Bsolanalytic}. This situation can be avoided, for example, if $0 \leq \alpha $ and $ 0 \leq \beta$, or if $\alpha\leq 0 $ and $ \beta <-1 $. The integration constant in \eqref{eq:phi0constraint} was conveniently named and parameterized by a constant called $\Lambda$. Furthermore, since our ansatz for the action \eqref{ActionI} already includes a linear term in $R$, one could simply set $\beta=0$, but we will keep it arbitrary for generality.  Condition \eqref{eq:phi0constraint} arises from solving the third order differential equation for $F$ obtained when making $\varphi = 0$. 

This condition is very restrictive since it does not allow all $F$ to have SSS metrics with a single degree of freedom. Actually, upon obtaining a solution for $R(r)$, this constraints the functional dependence of $F$. Not every Lagrangian will satisfy the constraint \eqref{eq:phi0constraint}. Fortunately, as mentioned before, this case is sufficiently simple so that we can obtain a closed formula for $R(r)$ and $B(r)$.

To find closed solutions, one replaces \eqref{eq:phi0constraint} into equation \eqref{eq:R} and into \eqref{eq:Bsolanalytic}, yielding an integro-differential equation for $R(r)$. To solve integro-differential equations, one can take derivatives of the equations until they become purely differential, solve the resulting differential equations with ordinary methods, and later fix the arbitrary constants such that the solution satisfies the original integro-differential equations. Following this procedure, we obtain a solution for $R(r)$, which in turn allowed us to find the function $B(r)$.  The result of such procedure is the function \eqref{eq:Bmetricphi0result}

\begin{widetext}

\be
B(r) =1-\frac{2
   M}{r}-\frac{\Lambda}{3}  r^2+\frac{3 \alpha  M}{(\beta +1)}-\frac{\alpha   (\beta +6 \alpha  M+1)}{(\beta +1)^3}\left(\alpha\log
   \left(\frac{r}{\beta +\alpha  r+1}\right)r^2+(\beta +1)r\right)\, ,
   \label{eq:Bmetricphi0result}
\ee
\end{widetext}
where $M$, $\alpha$, $\beta$ and $\Lambda$ are arbitrary constants. Of course, the obtained solution for $R(r)$ can be derived from \eqref{eq:Bmetricphi0result} by setting $\varphi$ and its derivatives to zero in \eqref{eq:ricciscalarfrommetricfunctions}.

From \eqref{eq:phi0constraint} and \eqref{eq:Bmetricphi0result}, one can notice that the Einstein-Hilbert case is recovered when $\alpha=0$, independently of the value of $\beta$. This is to be expected, since $\alpha$ different from $0$ is the only case where the functional constraint \eqref{eq:phi0constraint} allows the Lagrangian to differ from a simple linear function of the scalar curvature. 

A number of interesting properties come with this solution. To begin with, one can notice that near $r=0$, the metric behaves as
\be
B(r) \simeq -\frac{2M}{r} + \mathcal{O}(1) \, ,
\ee
and the scalar curvature as 
\be
R(r) \simeq -\frac{6 \alpha  M}{(\beta +1) r^2} + \mathcal{O}(\frac{1}{r}) \, ,
\ee
meaning that, with a single metric degree of freedom, the singularity at the center is unavoidable in any $F(R)$ theory in vacuum. \textit{In other words, regular SSS metrics with a single metric degree of freedom cannot come from a vacuum solution in $F(R)$ theories of gravity}. This is the strongest claim of our work. 

Furthermore, we observe that for $\alpha \neq 0$, the solution at large $r$ behaves as
\begin{align}
B(r) \simeq & \, r^2 \left(\frac{\alpha ^2 \log (\alpha^2 ) (\beta +6 \alpha  M+1)}{2(\beta +1)^3}-\frac{\Lambda }{3}\right) \nonumber \\
   & + \frac{1}{2} + \mathcal{O}(\frac{1}{r}) \, ,
\end{align}
which implies that at least in the asymptotic regimes $r \to 0$ and $r \to \infty$, the solution behaves as the Schwarzschild-(anti)de Sitter metric, differing only by proportionality factors. One can notice that the case where $\alpha \neq 0$ is only physically meaningful if $\frac{\alpha ^2 \log (\alpha^2 ) (\beta +6 \alpha  M+1)}{2(\beta +1)^3} \neq \frac{\Lambda }{3}$, otherwise the solutions approach the value $\frac{1}{2}$ at infinity, leading to a flat solution that violates Lorentz invariance. To align this solution with physical observations, one has to fine-tune the values of $\alpha$ and $\Lambda$ such that it accommodates observations. Only extremely small values of  $\alpha$ and $\Lambda$ would resemble known solutions in some finite range of $r$. 

\begin{figure}[h]
    \centering
    \includegraphics[width=\columnwidth]{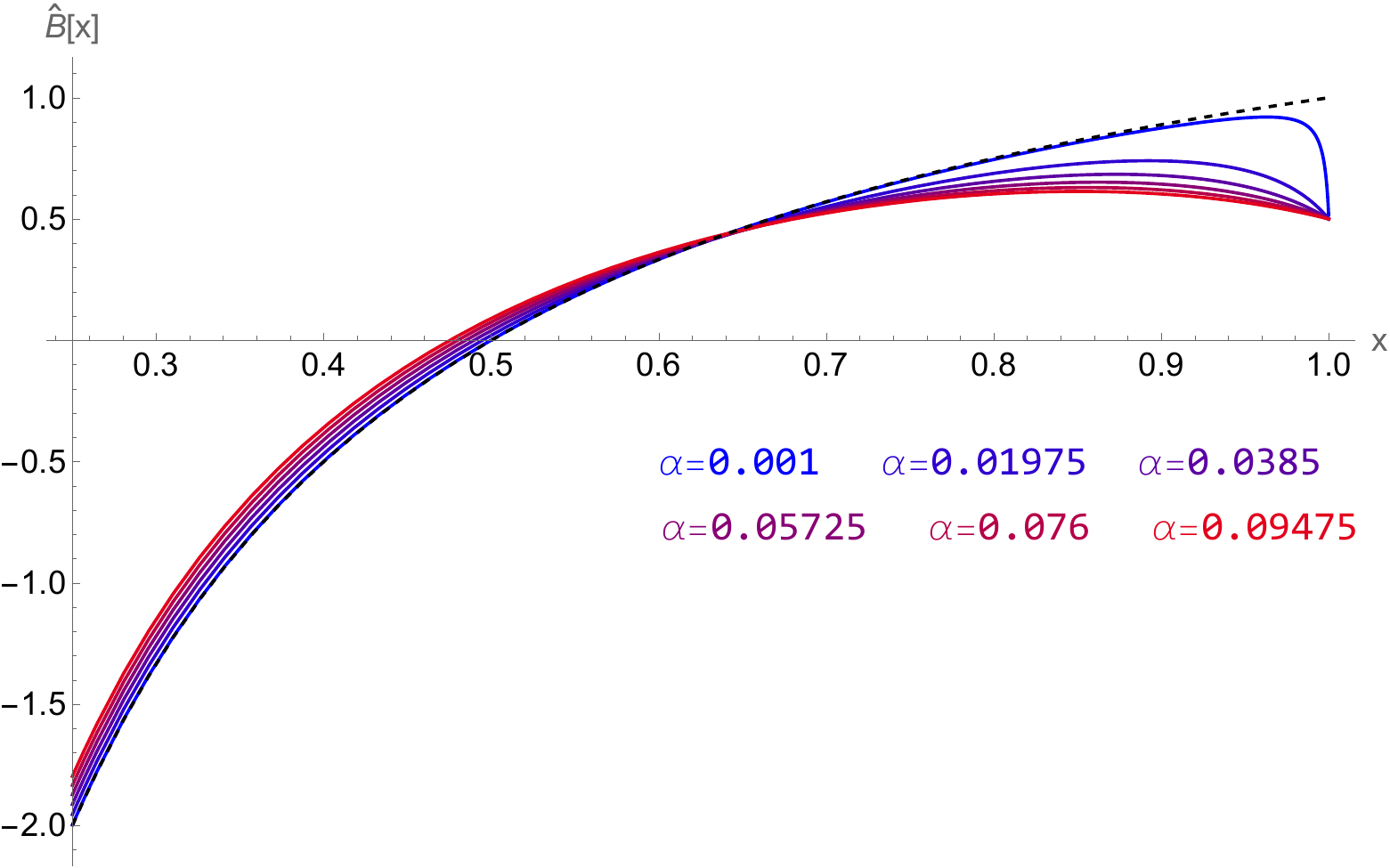}
    \caption{Examples of \eqref{eq:Bmetricphi0result}, as a function of the compact variable $x=\frac{\frac{r}{2M}}{1+\frac{r}{2M}}$, for $\Lambda=\frac{3 \alpha ^2 \log (\alpha^2 ) (\beta +6 \alpha  M+1)}{2(\beta +1)^3} $, $\beta=0$ and $M=1$ and different values of $\alpha$. The case of $\alpha=0$ is represented with a dashed line.}
    \label{fig:singlemetricexamples}
\end{figure}

On top of this, one can notice that these solutions provide a method for the reconstruction of the only allowed $F(R)$ Lagrangian compatible with a single metric degree of freedom. Indeed, fitting the curve $f(r) := \{ R(r) , F(R(r))   \}$ over a certain range of values of $r$ provides a way to reconstruct $F(R)$. Due to the nonphysical features of the obtained metric, we did not follow this program.

Our results show that if one desires to work in vacuum with a single metric degree of freedom within $F(R)$ gravitational theories, it is impossible to obtain solutions different from the Schwarzschild-(anti)de Sitter solution by freely choosing $F(R)$. In fact, in order to allow for such single metric solutions, the Lagrangian is constrained by \eqref{eq:phi0constraint}.

Furthermore, we found that the only possible solution generalizing the Schwarzschild-(anti)de Sitter solution exhibits nonphysical properties, such as failing to resemble asymptotically flat Minkowski space, unless the free parameters that differentiate the solution from Schwarzschild-(anti)de Sitter are set to zero. We conclude that our results do not favor $F(R)$ gravitational models in the case of a single metric degree of freedom.

In Fig. \ref{fig:singlemetricexamples}, one can see examples of asymptotically flat single metric solutions, where we made $\Lambda=\frac{3 \alpha ^2 \log (\alpha^2 ) (\beta +6 \alpha  M+1)}{2(\beta +1)^3} $, $\beta=0$ and $M=1$ for simplicity. The plots are performed using the compact variable $x \equiv \frac{\frac{r}{2M}}{1+\frac{r}{2M}}$, where we show the function $B(r)=B(\frac{2M x}{1-x})\equiv \hat{B}(x)$. This coordinate compactifies the spatial direction, such that the following mappings between spatial points hold: $r=\infty \to x=1$, $r=0 \to x=0$ and $r=2M \to x=\frac{1}{2}$, thus allowing us to show the asymptotic behavior of the obtained solution more clearly.

\section{ The case of $\varphi \neq 0 $ : \\
Approximation schemes}
\label{sect.approximations}
Having established that recovering Minkowski in the asymptotic regime is not possible with a single metric degree of freedom (unless one uses the Schwarzschild solution), we will now move beyond this limitation and study spacetimes with two metric degrees of freedom under certain approximations. This scenario complicates the resolution of \eqref{eq:R}, and therefore we must resort to some approximations. Despite the fact that one could explore this solution space by performing local expansions, such as the Frobenius method, this is not the approach we will follow in this section. Instead, we will explore approximations that have a chance of providing global information about the solutions, instead of just local information.

\subsection{\textit{Small $F$ approximation}}\label{subsect.lowF}

One can gain some intuition about the solutions found in \eqref{eq:Bsolanalytic} and \eqref{eq:phisolanalytic} by expanding them at small values of $F$ and its derivatives. This approximation is valid only when the corrections to GR are small. One can think of it as either assuming that the involved couplings in $F$ are small, making $F$ sub-leading compared to Einstein's action.  One can retain our expressions up to linear order in $F$ and its derivatives, giving
\begin{align}
B(r)\simeq &1-\frac{2M}{r}+\frac{\int r^2 F  \, dr}{2 r} \nonumber \\
& + \frac{r (2 M-r) R' F_{R} +M F_{R} }{r}, 
\label{eq:lowFBSol}
\end{align}
where the arbitrary constant $c_1$ was renamed $-2M$. The resulting equation for the scalar curvature is
\be
R(r)\simeq -2F  +\frac{3}{r^2}\left( (r-2M r^2) R' F_{R}   \right)'.
\label{eq:lowFREQ}
\ee

Despite not being necessary to obtain the previous expressions, one can also expand $\varphi$ to the lowest order in $F$, obtaining
\be
\varphi(r)\simeq \varphi_{0}+r R' F_{R} -F_{R} .
\label{eq:lowFPhiSol}
\ee

Using these expansions, one can clearly see how the results from Einstein-Hilbert gravity are recovered when $F=-2\Lambda$, recovering the results obtained in sec. \ref{sec:constantcurvature}.

It is worth noting in \eqref{eq:lowFBSol} that the first integral term corresponds to the cosmological constant in $B(r)$ when $F(R)$ is constant. As with the other corrections, one of them vanishes when $r=2M$, but might have a relevant impact at large values of $r$. The term proportional to $\frac{M}{r}$, could affect the effective mass of the system at smaller values of $r$.

Furthermore, an interesting feature of this linear approximation in $F$ is that the equation for the scalar curvature $R$ decouples from $B$. This allows for a simpler exploration of a vast amount of different functionals $F$ and their effect on the spacetime. 
Unfortunately, even in this simplified scenario, finding a solution for $R(r)$ without assuming a specific form for $F$ proved to be out of our reach. A numerical analysis would be required for each election of $F$, so to keep the generality of this paper, we leave this investigation for future work.

\smallskip


\subsection{\textit{Near extreme approximation}}\label{subsect.nearextreme}

We will now study another situation where we can make concrete statements about the spacetime without fixing to a particular $F(R)$. In order to obtain solutions that generalize the Schwarzschild-(anti)de Sitter case, we will now assume that the scalar curvature remains nearly constant, close to a minimum of the function $F(R)$. From \eqref{eq:constricciscalareq}, an extreme point of $F$ corresponds to the relation $2F(R_0)=-R_0$. Based on this, in this section we will name $R_0=-2\Lambda$, and we will approximate $F\simeq -2\Lambda + \frac{f_{2}}{2}(R-4\Lambda)^{2}$. Note that this is not constraining to a particular Lagrangian, since we are only assuming the existence of an extreme point of $F(R)$.

If the curvature remains close to this extreme of $F$, one must assume $R(r)=4\Lambda + \epsilon(r)$, with $|\epsilon(r)| \ll 1$. As discussed in sect.\ref{sect.3}, the case of $f_{2}<0$ is less physically interesting, since it cannot recover asymptotic flatness at infinity, and thus we will only focus on the case $0\leq f_{2}$.

Staying at linear order in $\epsilon$ and its derivatives, one obtains
\be\label{eq:vpapprox}
\varphi(r) \simeq \varphi_{0}+f_2 \left(r \epsilon'(r)-\epsilon(r)\right) \, ,
\ee
and
\begin{align}\label{eq:Bapprox}
B(r) \simeq & \, 1-\frac{2 M}{r}-\frac{\Lambda}{3} 
   r^2 + \nonumber \\ & \frac{f_2   \left(r
   \epsilon'(r) \left(6 M+\Lambda  r^3-3
   r\right)+\epsilon(r) \left(3 M-\Lambda 
   r^3\right)\right)}{3 r}
\end{align}
where $M$ is a free parameter. This expression matches with the approximation at low $F$ obtained in sec. \ref{sect.3}, when assuming that the product $f_2\, \Lambda$ is very small. 

The equation dictating the radial dependence of $\epsilon(r)$ comes from replacing \eqref{eq:vpapprox} and \eqref{eq:Bapprox} into the equation for $R$, \eqref{eq:R}, and is given by
\begin{align}
   r^2 \epsilon(r)(1 - 4 \Lambda f_2)= 3 f_2 \left(r  ( r-2 M -\frac{\Lambda}{3}r^3)\epsilon'(r)\right)' \, ,
   \label{eq:lineqcurvature}
\end{align}
at linear order in $\epsilon$ and its derivatives.

It can be observed that when $f_2 = 0$ the scalar curvature perturbation $\epsilon$ is forced to be zero, recovering the Schwarzschild-(anti)de Sitter solution.

One might also notice that at the roots of the equation $r  (r-2 M -\frac{\Lambda}{3}r^3)=0$, the differential equation has singular points, since the coefficient of the highest order derivative of $\epsilon$ vanishes. If the solutions present singularities at these points, then the linear approximation breaks down. Nevertheless, there could be solutions that avoid such singularities, making it worthwhile to study the linear perturbations both analytically and numerically. 

Before proceeding with a case-by-case study, one could notice that when one chooses $\Lambda=\frac{1}{4 f2}$, the equation can be solved analytically. Unfortunately, there is no non-trivial solution that remains finite at the roots of $r  (r-2 M -\frac{\Lambda}{3}r^3)=0$. 

With this in mind, we will now study the case of $M=0$ and $\Lambda \neq 0$ and the case of $M>0$ and $\Lambda = 0$, which are associated with perturbations of the Schwarzschild-(anti)de Sitter and the Schwarzschild metrics, respectively.

\subsubsection{The case of $ M = 0$}

In this section, we will assume that $M=0$ and $\Lambda \neq 0$. This means that we are analyzing perturbations around the de Sitter solution. In this case, one can solve the equation for the perturbation $\epsilon$ exactly, and obtain
\be
\epsilon(r)=\frac{c_1 \, P_{\gamma}\left(\frac{\sqrt{\Lambda }
   r}{\sqrt{3}}\right)+c_2 \, Q_{\gamma}\left(\frac{\sqrt{\Lambda } r}{\sqrt{3}}\right)}{r} \, ,
\ee
where $P_{\gamma}$ and $Q_{\gamma}$ are Legendre functions with $\gamma = \frac{1}{2} \left(\frac{\sqrt{25 f_2 \Lambda -4}}{\sqrt{f_2 \Lambda}}-1\right)$, and $c_1$, $c_2$ are arbitrary constants. Of course, we are only interested in the real part of this solution, and since the equation is linear in $\epsilon$, one can simply ignore the imaginary part of the solutions.

There are several interesting features that one can explore from these solutions.
   
For $\Lambda>0$, the Legendre function $Q_{\gamma}$ always has a divergence at $r=\sqrt{\frac{3}{\Lambda}}$, breaking the linear approximation at a finite radius. Furthermore, the $1/r$ factor introduces a potential divergence at $r=0$. While one could avoid one of those two divergences by appropriately choosing $c_1$ or $c_2$, the only way to avoid both divergences is to set $c_1=c_2=0$, which is just the de Sitter solution.
   
For the case of $\Lambda<0$, one still has the potential divergence at $r=0$, but there is no other divergence at a finite radius, so one could choose $c_1$ or $c_2$ such that this divergence is removed. In this scenario, it is possible to fine tune the extra free parameter such that $\epsilon(r)$ is arbitrarily small for all finite values of $r$. Nevertheless, eliminating the divergence at $r=0$, one still has the problem that at $r \to \infty$, $|\epsilon|$ tends to infinity. The only way to remove the divergence at $r=0$ and force the solution to be finite when $r \to \infty$ is by setting $c_1=c_2=0$, again recovering just the anti-de Sitter solution.

One can think of these results as an instability of the de-Sitter ($\Lambda>0$) solution in $F(R)$, since small deviations from it eventually diverge.

On the contrary, in the case of an anti-de Sitter solution ($\Lambda < 0$), perturbations can be made arbitrarily small for all finite values of $r$, meaning that perturbations of anti-de Sitter remain stable under the inclusion of a non-trivial $F(R)$.

\subsubsection{The case of $\Lambda = 0$}
In this section, we will use $\Lambda = 0$ and $M>0$.
Unfortunately, in this case, we could not solve \eqref{eq:lineqcurvature} analytically. Nevertheless, one can check that asymptotic behavior of $\epsilon(r)$ in the limit of $r \to \infty$ is
\be
 \epsilon(r) \simeq \frac{1}{r}(c_1 e^{\frac{- \, r}{\sqrt{3 f_2}}}+c_2 e^{ \frac{ \,r}{\sqrt{3 f_2}}}) + \mathcal{O}(\frac{1}{r^2}) \, ,
 \label{eq:asymptoticcurvaturelabdazero}
\ee
where $c_1$ and $c_2$ are arbitrary constants, matching with the first order in perturbations solution found in \eqref{eq:asymptoticsolRflat}. 

\begin{figure}[h]
    \centering
    \includegraphics[width=\columnwidth]{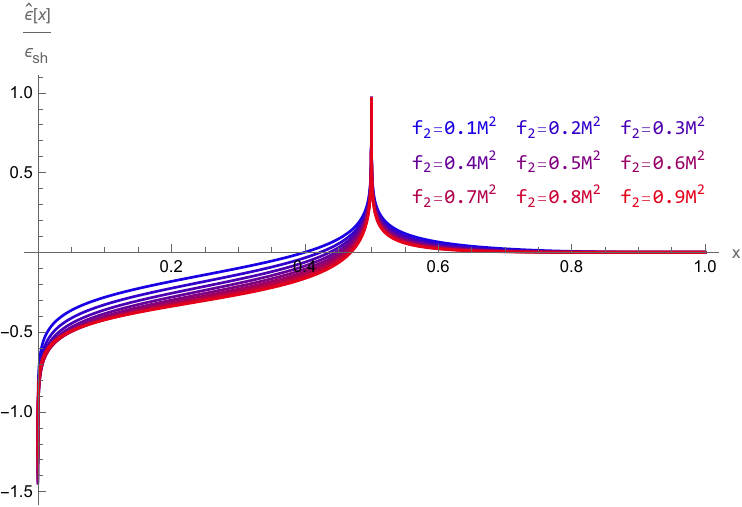}
    \caption{Examples of the numerical solutions for $\hat{\epsilon}(x)$, for different values of the ratio $\frac{M^{2}}{f_{2}}$. The plot is normalized by the value of $\hat{\epsilon}$ at $r=2M$ ($\epsilon_{sh}$). One can observe the cusp at the finite radius $r=2M$, which breaks the linear approximation.}
    \label{fig:epsiloncorrectionminimum}
\end{figure}

To obtain a full description of $\epsilon$ for all values of $r$ in this approximation, we resort to numerical methods. Using the variable $x=\frac{\frac{r}{2M}}{1+\frac{r}{2M}}$ defined in previous chapters, one can solve for $\epsilon$ numerically more easily than using the $r$ variable. In terms of this new variable, one can define $\epsilon(r)=\epsilon(\frac{2M x}{1-x})\equiv \hat{\epsilon}(x)$. After this change of variables, eq. \eqref{eq:lineqcurvature} with $\Lambda=0$ takes the form 
\be
x^2 \hat{\epsilon}(x)=
   \frac{3 f_2}{4 M^2}(x-1)^4\left(x (2 x-1) \hat{\epsilon}'(x)\right)' \, ,
   \label{eq:lineqcurvaturexvariable}
\ee
where now the only free parameter in these coordinates is the ratio $p=\frac{M^{2}}{f_{2}}$.

In order to obtain numerical solutions, one has to fix at least two boundary conditions. Fixing the value of $\hat{\epsilon}$ at a point and its first derivative proved to be non-fruitful, since independently if this was done at the origin $r=0$ ($x=0$), at $r=2M$ ($x=\frac{1}{2}$) or at infinity $r=\infty$ ($x=1$), the numerical solver always encountered divergences at some other finite value of $x$. Due to this, we decided to put the constraints that $\hat{\epsilon}$ should be $0$ at infinity, recovering Minkowski space, and take a finite value at $x=0$. Since the equation is linear and homogeneous in $\epsilon$, this value is arbitrary. Using these boundary conditions, finite element methods allow us to obtain solutions across the entire domain.

In Fig. \ref{fig:epsiloncorrectionminimum}, we show the numerical results for $\hat{\epsilon}(x)$ for the different values of $p$. These results are normalized by the value of the curvature at $r=2M$, denoted as $\hat{\epsilon}(x=\frac{1}{2})\equiv\epsilon_{sh}$, which is a free parameter due to the linear nature of eq. \eqref{eq:lineqcurvaturexvariable}.

These results indicate that while the curvature remains finite at all values of $r$, it exhibits a cusp at $r=0$ and also at $r=2M$, thus breaking the linear approximation. Our results suggest that it is impossible to have small corrections to the Schwarzschild solution in the case where $F(R)$ is near a quadratic extreme. The assumption that corrections could remain small up to the horizon proved to be inconsistent, and therefore we find this result non-trivial. Our results are limited to the case of a quadratic minimum, and they do not generalize to cases where the extreme is not quadratic, but cubic, quartic, or of any other order. We leave the analysis of different extremes of $F(R)$ for future work.

As in the case of $M=0$, these results can be viewed as an instability of the Schwarzschild solution. Deviations that start infinitesimally close to the Schwarzschild solution at infinity, will eventually diverge. From this perspective, one could conclude that these results do not favor $F(R)$ gravitational models, since favorable results should indicate that small deviations from well-understood solutions, such as de-Sitter or Schwarzschild, remain small. However, there is another way to interpret these findings. One could argue that these results suggest that Schwarzschild and de Sitter are not robust gravitational solutions, since any slight modification to the gravitational action makes them unstable.


\section{Axial gravitational waves}\label{sec:gravgraves}
\footnote{We thank Badri Krishnan for his suggestion on performing the calculations that are presented in this section.}In this section, we compute axial perturbations of the metric \eqref{eq:ansatzmetric}, in the Regge-Wheeler gauge \cite{Regge:1957td}. Our attention will be focused on deriving the potential of odd parity waves modes. Using the Regge-Wheeler gauge, such perturbations of the metric take the form 

\begin{equation}
h_{\mu \nu} = e^{-i \omega t} \sin \theta \, \frac{\partial}{\partial \theta} P_l(\cos \theta) 
\begin{pmatrix}
    0 & 0 & 0 & h_0(r) \\
    0 & 0 & 0 & h_1(r) \\
    0 & 0 & 0 & 0 \\
    h_0(r) & h_1(r) & 0 & 0 
\end{pmatrix} \, ,
\label{eq:axialpertmetric}
\end{equation}
with $w$ the frequency, $P_l$ a Legendre polynomial, and $h_0$ and $h_1$ radial perturbation functions. 

We will generalize the procedure in \cite{Regge:1957td} to derive the equations of motion of $h_0$ and $h_1$. Introducing \eqref{eq:axialpertmetric} into \eqref{eq:vacuumeqfr}, and retaining only first-order terms in $h_{\mu \nu}$, one can notice that there are only two independent equations, namely the $r \phi$ and $\theta \phi$-components of the perturbed vacuum equations \eqref{eq:vacuumeqfr}. Due to their length, these equations are presented in the appendix, see \eqref{eq:perturbedrphi} and \eqref{eq:perturbedphiphi}. Notably, the Ricci scalar remains unchanged to first order in \eqref{eq:axialpertmetric}, which greatly simplifies the analysis of the perturbed $F(R)$ vacuum equations of motion.

From \eqref{eq:perturbedphiphi} one can determine $h_0$ in terms of $h_1$, and replace it into \eqref{eq:perturbedrphi}, obtaining a second order equation for $h_1$. This equation can be put in its normal form by performing the variable change 
\be
h_1(r)=\frac{r \, Q(r) }{B(r)}\left(\frac{e^{-\frac{\varphi (r)}{2}}}{\sqrt{1+F_R(R(r)) }}\right) \, ,
\label{eq:vartransfnormal}
\ee
with $Q(r)$ being an unknown radial function. 

In the original work by Regge and Wheeler, the transformation \eqref{eq:vartransfnormal} did not include the factor in brackets, which is equal to $1$ for their case since they perturbed the Schwarzschild solution using Einstein's action. In our more general case, this extra factor is needed, otherwise the equation for $Q$ is not in normal form. The transformed equation becomes
\be
\frac{\partial^2 Q(r)}{\partial r_{*}^2} +Q(r)(w^2 - V_{eff}(r))=0 \,\,\,\,\,\,\,\,\,\,\,\,\,\,\,\, \frac{dr_{*}}{dr}\equiv \frac{e^{-\frac{\varphi (r)}{2}}}{B(r)} \, , 
\ee
where we defined the tortoise coordinate, denoted as $r_{*}$, the potential $V_{eff}$ in \eqref{eq:effectivepotQN}. All quantities in this expression are evaluated at $r$.

One should substitute our found solutions \eqref{eq:phisolanalytic} and \eqref{eq:Bsolanalytic} into \eqref{eq:effectivepotQN} to use the full extent of our formalism. However, due to the complexity of the resulting expression in terms of the Lagrangian $F$ and the scalar curvature, we decided to present the result without this replacement. Note that \eqref{eq:effectivepotQN} is not directly the potential derived from the perturbed vacuum equations of motion \eqref{eq:perturbedrphi} and \eqref{eq:perturbedphiphi}; rather, we simplified it by using \eqref{eq:R}, \eqref{eq:B} and \eqref{eq:varphi} to avoid higher-order derivatives of the metric functions and the scalar curvature. 

\begin{widetext}
    \be
   V_{eff}(r) \equiv \frac{1}{4} B e^{\varphi} \left(\frac{4 \left(l^2+l-3\right)}{r^2}+\frac{3 B \left(2+r R' F_{RR}+2
   F_R\right)^2}{r^2 \left(1+F_R\right)^2}-\frac{2
   (F+R)}{(1+F_R)}+2 R\right) \, .
   \label{eq:effectivepotQN}
    \ee
\end{widetext}

Setting $F=R=\varphi=0$ and their derivatives to zero, along with using the Schwarzschild solution for $B$ in \eqref{eq:effectivepotQN}, we recover the classical result of Regge and Wheeler \cite{Regge:1957td}. 

As before, from the obtained exact expressions it is difficult to gain some intuition on the impact of the extended Lagrangian. Nevertheless, in section \ref{sect.approximations} we introduced the approximation at low $F$, where only the first-order terms in the Lagrangian modification $F$ are retained. Substituting the expansions \eqref{eq:lowFBSol}, \eqref{eq:lowFREQ} and \eqref{eq:lowFPhiSol} into the effective potential, we obtain the first-order expansion in $F$ for the effective potential \eqref{eq:LowFVeff}
\begin{align}
    V_{eff}(r) & \simeq  \frac{ (r-2 M) (l (l+1) r-6 M) e^{\varphi_0}}{r^4} \nonumber \\ 
    & - \Bigg[ 2 F_R \left(24 M^2-3 \left(l^2 + l + 3\right) M r 
        + l (l + 1) r^2 \right)  \notag \\
    & \qquad + \left(12 M - \left(l^2 + l + 3\right) r\right) 
        (\int r^2 F \, dr ) \notag \\
    & \qquad + r^3 (r - 2 M) F \Bigg]\frac{e^{\varphi_0}}{2 r^4} \, ,
    \label{eq:LowFVeff}
\end{align}
where all quantities in \eqref{eq:LowFVeff} are evaluated at $r$, which we avoided for notational simplicity. Here, $\varphi_0$ is kept arbitrary for generality.

As expected, the first line of \eqref{eq:LowFVeff} is precisely the Regge-Wheeler potential, while the subsequent terms introduce first-order corrections due to the low value of the Lagrangian extension $F$. The second line introduces corrections that are likely to be more relevant at small distances, and the last two lines are the ones that could have a bigger impact at large spatial distances. 

These results pave the ground for exploring observational effects of $F(R)$ gravitational theories within black hole perturbation theory. In particular, due to the duality at the classical level of $F(R)$ theories with scalar-tensor theories, our work can be directly mapped to non-vacuum GR, including scenarios with scalar matter.



\section{Summary and Outlook}\label{sect.Concl}

In this paper, we reduced the complexity of the vacuum equations of motion for static and spherically symmetric spacetimes, from two coupled fourth-order differential equations for the metric functions, to a single second-order equation for the scalar curvature of spacetime. For the first time, we derive analytic expressions for the metric functions in terms of an arbitrary gravitational theory $F(R)$, providing direct insight into how the Lagrangian impacts the spacetime. From our simplified equations, we demonstrated that the solution space is parameterized by at most four arbitrary constants, considerably reducing the arbitrariness one might expect by directly looking at the equations of motion for the metric functions. 

Using the metric solutions and reduced equations of motion, we studied the conditions for asymptotic flatness, the existence of event horizons, and regularity at the core. We found that only Lagrangians with $0 \leq F_{R}(0)$ can recover Minkowski space at spatial infinity, when the solution is not exactly the Schwarzschild solution. Moreover, we found that by fixing some of the arbitrary constants determining the solution space, it is possible to impose the existence of solutions with an event horizon. Furthermore, we showed that under certain conditions, it is also possible to impose, simultaneously with the existence of an event horizon, regularity at the core. This situation exhausts the amount of free constants to fix, leaving the asymptotic behavior of the solutions to depend on the specific $F(R)$ theory. 

For solutions with an event horizon, we studied their thermodynamical properties such as entropy, temperature, and specific heat. To the best of our knowledge, this is the first time that formulas for the temperature and specific heat have been derived in terms of an arbitrary $F(R)$. From our formula for the temperature, we showed that it is possible to have extremal black holes under certain parameter choices. Furthermore, we noticed that it is theoretically possible to achieve simultaneously zero temperature and entropy, fulfilling the third law of thermodynamics, although such limit must carefully be handled since it might give rise to singularities. 

We also addressed the problem of generalizing the Schwarzschild-(anti)de Sitter solution in the case of a single metric degree of freedom. We found the most general static and spherically symmetric spacetime solution in this context. However, the case of a single metric degree of freedom imposes severe constraints on possible gravitational theories $F(R)$. Even though the found solution extends Schwarzschild-(anti)de Sitter solution with two arbitrary parameters, it possesses some unphysical features. For example, it does not recover asymptotic flatness unless the extra parameters are chosen such that the solution is equal to the Schwarzschild solution. Also, new singularities appear for some parameter elections. Based on these findings, we argue that our results do not favor $F(R)$ gravitational models in this situation. 

For the more general case of two metric degrees of freedom, we discuss approximation schemes to find solutions beyond the Schwarzschild-(anti)de Sitter solution. In particular, we studied the situation where small curvature corrections to this solution exist throughout the spacetime. Assuming that the $F(R)$ model can be expanded near a minimum, we carried out both analytical and numerical studies to check whether our assumption of small curvature corrections would hold. In all the analyzed cases, the corrections assumed small were found to diverge at a finite value of the radial coordinate $r$, with the exception of the anti-de Sitter solution. Our results imply that small scalar curvature corrections to the GR solutions will eventually diverge. 

There are two ways to interpret these results. From an effective field theory perspective, where higher curvature corrections to GR are expected to appear in the action, one could interpret these results as an instability of the Schwarzschild-de Sitter solution under the inclusion of a Lagrangian extension $F(R)$, meaning that Schwarzschild-de Sitter is a very particular solution that would not hold on a larger setting. On the other hand, if one views GR as a complete description of spacetime, these results could be seen as discouraging the adoption of $F(R)$ gravitational models, since the well understood Schwarzschild-de Sitter solution is unstable under this extension.

Following the lines of a stability analysis of static and spherically symmetric spacetimes, we studied the dynamics of axial perturbations of the vacuum solutions. Using the obtained analytical expressions for the metric functions, we derive the effective potential for quasinormal modes in terms of an arbitrary Lagrangian $F(R)$. The found potential reproduces the well-known Regge-Wheeler potential \cite{Regge:1957td} in the case of Schwarzschild solution and Einstein's action. We explore the expansion of the generalized potentials in powers of $F(R)$, assuming that corrections to GR are small. In this scenario, we identify the leading corrections, and which ones are likely to have a larger impact at small and large distance scales. Our results pave the way to investigate the observational impact of extended gravitational theories. 

In the second part of this work \cite{secondpartpapaer}, we will dive into exploring analytically and numerically the most commonly used $F(R)$ cosmological models, using our framework. Our goal will be to reduce the existing over-fitting of Cosmic Microwave Background observations with extended gravitational models by identifying which models allow for black hole solutions with well-defined asymptotic limits. We will also explore the existence of both extremal and non-extremal black holes, and we will attempt to simultaneously impose regularity of the solutions at the core. 

Investigating the consequences of our framework for treating $F(R)$ extensions to GR in the context of gravitational collapse \cite{Oppenheimer:1939ue} is an important next step. In particular, a review of the junction functions \cite{Israel:1966rt,Deruelle:2007pt,Senovilla:2013vra}, along with the well-known stringent constraints they impose on $F(R)$ gravity \cite{Cembranos:2012fd,Casado-Turrion:2022xkl}, would be necessary to better understand the parameter space. Achieving a realistic description of vacuum solutions and comparing models that successfully describe cosmological observations would also require the inclusion of angular momentum, which, while challenging, remains crucial for the completeness of these models. We aim to explore these directions in future works.

\section*{Acknowledgments} 

We would like to thank Renate Loll, Badri Krishnan, Frank Saueressig and Jesse Daas for enlightening conversations and useful feedback. The work
of C.L. is supported by the scholarship Becas Chile
ANID-PCHA/2020-7221007.

\bibliography{bibliography}

\begin{widetext}
\section*{Appendix}\label{sec:appendix}

\subsection{Vacuum equations of motion} 

For completeness, we include the tt and rr-components of the vacuum equations of motion for static and spherically symmetric spacetimes \eqref{eq:vacuumeqfr}. As mentioned in the body of the paper, these equations are non-linear, fourth-order, coupled differential equations on the metric functions $(B, \, \varphi)$. In these equations, one should replace the argument of $F$ and its derivatives by \eqref{eq:ricciscalarfrommetricfunctions}. 
\begin{align}
    &4 B^3 F_{RRR} \left(r^3 \varphi ^{(3)}+\left(r \varphi '+2\right) \left(r^2 \varphi
   ''-2\right)\right)^2 \nonumber\\
   &- 4 B^2 \Big(r^2 F_{RR} \left(r^3 \left(r \left(\varphi
   ^{(4)}+\left(\varphi ''\right)^2\right)+\varphi ' \left(r \varphi ^{(3)}+2 \varphi ''\right)+4 \varphi
   ^{(3)}\right)+4\right) \nonumber\\
   &\qquad -F_{RRR} \left(r^3 \varphi ^{(3)}+\left(r \varphi '+2\right)
   \left(r^2 \varphi ''-2\right)\right) \left(r B' \left(r \left(5 r \varphi ''+\varphi
   ' \left(r \varphi '+4\right)\right)-4\right)+r^2 (2 B^{(3)} r+B'' \left(3 r
   \varphi '+8\right))+8\right)\Big) \nonumber\\
   &- r^3 \left(2 r^3 B'' F_R+r \left(B'\right)^2
   F_{RR} \left(r \left(5 r \varphi ''+\varphi ' \left(r \varphi
   '+4\right)\right)-4\right)+B' \left(F_{RR} \left(r^2 \left(2 B^{(3)} r+B'' \left(3 r
   \varphi '+8\right)\right)+8\right)\right.\right.\nonumber\\
   &\qquad \left.\left. + r^2 F_R \left(3 r \varphi '+4\right)-4 r^2\right)+2
   r^3 F+4 r\right) \nonumber\\
   &+ B \Big(F_{RRR} \left(r B' \left(r \left(5 r
   \varphi ''+\varphi ' \left(r \varphi '+4\right)\right)-4\right)+r^2 \left(2 B^{(3)} r+B''
   \left(3 r \varphi '+8\right)\right)+8\right)^2 \nonumber\\
   &\qquad - 2 r^2 F_{RR} \left(r \left(B'
   \left(r \left(2 r \left(\varphi '\right)^2+4 r \left(2 r \varphi ^{(3)}+5 \varphi
   ''\right)+\varphi ' \left(5 r^2 \varphi ''-2\right)\right)-12\right)\right.\right.\nonumber\\
   &\qquad \left.\left. + r \left(B'' \left(r
   \left(8 r \varphi ''+\varphi ' \left(r \varphi '+10\right)\right)+4\right)+r \left(2 B^{(4)}
   r+3 B^{(3)} \left(r \varphi '+4\right)\right)\right)\right)-8\right)\nonumber\\
   &\qquad - \left(r^5 F_R \left(2
   r \varphi ''+\varphi ' \left(r \varphi '+4\right)\right)\right)+4 r^4\Big)=0 \, ,
   \label{eq:tteqvacuum}
\end{align}

\begin{align}
    &2 B^2 F_{RR} \left(r \varphi '+4\right) \left(r^3 \varphi ^{(3)}+\left(r \varphi '+2\right) \left(r^2 \varphi
   ''-2\right)\right) \nonumber\\
   &+ r \Big(2 r^3 B'' F_R + r \left(B'\right)^2 F_{RR} \left(r \left(5 r \varphi
   ''+\varphi ' \left(r \varphi '+4\right)\right)-4\right) \nonumber\\
   &\qquad + B' \left(F_{RR} \left(r^2 \left(2 B^{(3)} r+B'' \left(3 r
   \varphi '+8\right)\right)+8\right) + r^2 F_R \left(3 r \varphi '+4\right)-4 r^2\right) \nonumber\\
   &\qquad + 2 r^3 F+4 r\Big) \nonumber\\
   &+ B \Big(r^4 B' \left(\varphi '\right)^3 F_{RR}+r^3 \left(\varphi '\right)^2 \left(F_{RR} \left(3 r B''+8
   B'\right)+r F_R\right) \nonumber\\
   &\qquad + 2 \Big(F_{RR} \left(r \left(B' \left(r^3 \varphi ^{(3)}+12 r^2 \varphi ''-12\right)+4
   r \left(B^{(3)} r+4 B''\right)\right)+16\right)+r^4 F_R \varphi ''-2 r^2\Big) \nonumber\\
   &\qquad + r \varphi '
   \Big(F_{RR} \left(r \left(2 B^{(3)} r^2+20 r B''+B' \left(7 r^2 \varphi ''+8\right)\right)+8\right)-4
   r^2\Big)\Big)=0 \, .
   \label{eq:rreqvacuum}
\end{align}

\subsection{Axial perturbations of spacetime}\label{AppendixB}
As stated in the body of this work, in this part of the appendix we present the $r \phi$ and $\theta \phi$-components of the perturbed vacuum equations \eqref{eq:vacuumeqfr}. 
\begin{align}
    & r w \left(F_{R} + 1\right) \left(-i r h_{0}' + 2 i h_{0} 
        + r w h_{1} \right) - B h_{1} e^{\varphi} \Big( l (l+1) \left(F_{R} + 1\right) 
        - r \left(2 B' \left(-r R' F_{RR} + F_{R} + 1\right) 
        + r F + r R \right) \Big) \notag \\
    & + B^2 h_{1} e^{\varphi} \Big( \left(r \varphi' + 2\right) F_{R} 
        + r \Big( -2 r F_{RRR} R'^2 - F_{RR} \left(2 r R'' 
        + R' \left(r \varphi' + 2\right)\right) \Big) 
        + r \varphi' + 2 \Big)  = 0 \, ,
        \label{eq:perturbedrphi}
\end{align}
\begin{align}
    -B  e^{\varphi  } \Big(\left(F_{R} +1\right)
   \left( h_{1}  \left(2 B' +B  \varphi
   ' \right)+2 B   h_{1}' \right)+2 B 
    h_{1}  R'  F_{RR} \Big)-2 i w
    h_{0}  \left(F_{R} +1\right)=0
    \label{eq:perturbedphiphi} \, .
\end{align}

From \eqref{eq:perturbedphiphi} one can solve $h_0$ in terms of $h_1$, and replace it in \eqref{eq:perturbedrphi} from where one can finally obtain the result \eqref{eq:effectivepotQN}.

\end{widetext}

\end{document}